\documentclass[12pt]{article}
\usepackage{authblk}
\usepackage{natbib} 
\usepackage{graphicx}
\usepackage{amssymb} 
\usepackage{verbatim} 
\usepackage{tabularx}
\newcommand{\water}{H$_2$O}
\newcommand{\methanol}{CH$_3$OH}
\newcommand{\ammonia}{NH$_3$}
\newcommand{\formaldehyde}{H$_2$CO}

\newcommand{\kms}{km~s$^{-1}$}
\newcommand{\cc}{cm$^{-3}$}

\newcommand{\apj}{ApJ}
\newcommand{\apjs}{ApJS}
\newcommand{\apjl}{ApJL}
\newcommand{\mnras}{MNRAS}

\newcommand{\aap}{A\&A}
\newcommand{\pasj}{PASJ}
\newcommand{\nat}{Nature}

\title{{\bf Astro 2020 White Paper:} \protect\\ Understanding Accretion Outbursts in Massive Protostars through Maser Imaging \protect\\ {\normalsize {\bf Thematic Area:}  Star and Planet Formation}\vspace{-5mm}}
\author[ \hspace{-1ex}]{Todd Hunter$^1$ (520 Edgemont Rd, Charlottesville, VA, thunter@nrao.edu)}
\author[2]{\protect\\Co-authors: Anna Bartkiewicz}
\author[1]{Walter Brisken}
\author[1]{Crystal L. Brogan}
\author[3]{Ross Burns}
\author[4]{James O. Chibueze}  
\author[5]{Claudia J. Cyganowski}
\author[3]{Tomoya Hirota}
\author[6,7]{Gordon MacLeod}
\author[8]{Alberto Sanna}
\author[9]{Jos\'e-Mar\'{\i}a Torrelles\protect\vspace{-3mm}}
\affil[1]{\normalsize National Radio Astronomy Observatory (NRAO)}
\affil[2]{\normalsize Nicolaus Copernicus University, Torun, Poland}
\affil[3]{\normalsize National Astronomical Observatory of Japan (NAOJ)}
\affil[4]{\normalsize Centre for Space Research, North-West University}
\affil[5]{School of Physics and Astronomy, University of St. Andrews} 
\affil[6]{\normalsize Hartebeesthoek Radio Astronomy Observatory (HartRAO)}
\affil[7]{\normalsize University of Western Ontario}
\affil[8]{\normalsize Max-Planck-Institut f\"ur Radioastronomie (MPIfR)}
\affil[9]{\normalsize Institut de Ci\`encies de l'Espai (CSIC)}
\date{February 22, 2019}

\begin{document}
\maketitle
\thispagestyle{empty}
\vspace{-9mm}
\begin{abstract}
The bright maser emission produced by several molecular species at centimeter to long millimeter wavelengths provides an essential tool for understanding the process of massive star formation.
Unimpeded by the high dust optical depths that affect shorter wavelength observations, the high brightness temperature of these emission lines offers a way to resolve accretion and outflow motions down to scales below $\sim$1 au in deeply embedded Galactic star-forming regions at kiloparsec distances.
The recent identification of extraordinary accretion outbursts in two high-mass protostars, both of which were heralded by maser flares, has rapidly impacted the traditional view of massive protostellar evolution, leading to new hydrodynamic simulations that can produce such episodic outbursts.  In order to understand how these massive protostars evolve in response to such events, larger, more sensitive ground-based centimeter wavelength interferometers are needed that can simultaneously image multiple maser species in the molecular gas along with faint continuum from the central ionized gas.  Fiducial observations of a large sample of massive protostars will be essential in order to pinpoint the progenitors of future accretion outbursts, and to quantify the outburst-induced changes in their protostellar photospheres and outflow and accretion structures.
Knowledge gained from these studies will have broader impact on the general topic of accretion onto massive objects.
  
\end{abstract}
\clearpage
\setcounter{page}{1}
\section{The Importance of Masers in Star Formation}
\vspace{-3mm}
The process of star formation leads to the concentration of molecular gas to high densities in molecular cloud cores.  The potential energy released by gravitational collapse and accretion onto the central protostars heats and excites the surrounding material through infrared radiation and high velocity bipolar outflows.  Both of these feedback mechanisms (radiative and mechanical) naturally produce population inversions between specific pairs of energy levels in several abundant molecules, including \water, \methanol, OH, \ammonia, SiO, and \formaldehyde.  The resulting non-thermal maser emission in the corresponding spectral transitions provides a beacon whose brightness temperature far exceeds that of the more commonly-excited thermal emission lines. Consequently, maser lines at centimeter wavelengths have traditionally provided a powerful probe of star formation. 
In general, they trace hot, dense molecular gas, revealing the kinematics of star-forming material within a few 1000\,au of very young stars, including accretion disks and their associated jets, as well as shocks  where the jets impact ambient gas in the outflow lobes. Masers are generally more prevalent in regions surrounding massive protostars, due to their higher luminosities and more energetic outflows. Furthermore, masers are
sensitive indicators of sudden changes in the physical conditions near the protostars.  Recently, it has been recognized that maser flares in lines that are radiatively pumped
by infrared photons can be directly associated with bursts of accretion onto the central protostars.  In this context, maser emission provides a unique tool for probing how massive protostars accrete matter, allowing us to reconstruct the gas dynamics in their vicinity, as well as to study the accretion process in the time domain.

%
\vspace{-5mm}
\section{State of the Art and Current Limitations}
\vspace{-3mm}
With the advent of the Atacama Large Millimeter/submillimeter Array (ALMA), imaging weak thermal lines at high angular resolution has become feasible,
and recent results have begun to place previous and ongoing maser studies into better physical context \citep[see e.g., Orion Source I, $d\sim 420$~pc;][]{Plambeck16,Hirota17}.
At the distances of more typical massive star-forming regions ($d> 1$~kpc), however, the brightness temperature sensitivity of ALMA is still not sufficient
to trace the accretion flow and accompanying jet structures that surround massive protostars, because of the high angular and spectral resolution required.
Moreover, at the short wavelengths of ALMA, the combination of molecular line confusion and high dust opacity toward the hot cores will often hamper the direct imaging of accretion processes
close to the protostars.
In contrast, the centimeter maser transitions propagate unobscured from the innermost
regions, providing a strong signal for self-calibration, and thus enabling high dynamic range imaging on long baselines. 

Unfortunately, the angular resolution of the Very Large Array (VLA) is insufficient to study the details of accreting gas, particularly in the 6 GHz band where the resolution
is limited to $\sim$0.3$''$. 
In the handful of nearest examples of massive star formation (d$\sim$1~kpc), this resolution corresponds to 300~au.
However, in the majority of massive star-forming sites across the Milky Way located at several kpc from the Sun, it exceeds 1000~au, which is a problem because each site typically contains a cluster of massive protostars, a phenomenon that is often termed a ``proto-Trapezium'' \citep{Megeath05} or a ``protocluster'' \citep{Minier05}.  The  separation of protostars in these protoclusters is often $\lesssim$1000~au \citep[e.g.,][]{Palau13}.
Thus, to avoid source confusion, and ultimately to resolve the spatial morphology and kinematics
of disks or other accretion structures at scales of 1-10 au, requires
an improvement in angular resolution of 2-3 orders of magnitude.
Such a resolution would also enable three-dimensional
measurements of gas velocity via multi-epoch proper motions. 

While current VLBI facilities (VLBA, EVN, eMERLIN, KVN, VERA, and LBA) have the requisite angular resolution to detect maser proper motions,
studies at these scales currently suffer from poor surface brightness sensitivity and uv coverage (giving a relatively low dynamic range) that limit detections to non-thermal processes exceeding brightness temperatures of $T_B\sim10^7$\,K \citep[e.g.][]{Matsumoto14,Bartkiewicz09}. This limited sensitivity hinders the science in two key ways.  First, only the brightest maser
spots can be detected, reducing the fidelity with which kinematic structures can be delineated in a single epoch, and reducing the number of spots that will potentially
persist over multiple epochs (used for proper motion studies).   While the current VLBI capability is sufficient for measuring the bulk proper motion of a star-forming region, as in the Bar and Spiral Structure Legacy Survey 
\citep[BeSSeL; e.g.][]{Reid2014}, it is insufficient to disentangle the gas kinematics surrounding each protostar within a young cluster.  Second, the thermal radio continuum emission ($T_B \sim 10^4$~K) that arises in the immediate vicinity of very young massive protostars, with typical flux densities of $<1$\,mJy \citep{Cyganowski2011,Rosero2016,Moscadelli2016,Sanna18}, cannot be observed simultaneously with the masers, leading to (relative) positional uncertainties between the protostellar and maser components.  The resulting ambiguity of the dynamical center severely hinders the interpretation of multi-epoch measurements, which are essential to understand the mass, momentum, and kinetic energy of the inner jet where it transitions into a bipolar molecular outflow.  Studying these objects at scales of 1-10~au in a comprehensive list of maser lines, and with sufficient sensitivity to image simultaneously
the associated continuum emission (either thermal or synchrotron) on the shorter baselines, is an essential goal for future facilities. 

\vspace{-7mm}
\section{Outburst phenomena probed by maser transitions}
\vspace{-4.5mm}
	\label{impact}
	
The massive protostars in protoclusters usually exist in diverse evolutionary states \citep[e.g.,][]{Brogan2016}, and exhibit emission in different maser lines, each offering a unique view into particular phenomena of massive star formation \citep{Menten2007}. The Class II \methanol\/ maser lines, primarily at 6.7~GHz, 12.2~GHz, and 19.9~GHz, trace hot molecular gas that is close  ($\lesssim$ 1000~au)
to the youngest massive protostars, which can provide the intense mid-infrared emission \citep[e.g.,][]{Moscadelli2011,Bartkiewicz14} required to 
pump the maser transitions \citep{Sobolev97,Cragg05}.  The light curves of this maser species show intriguing variations that are likely caused by changes in the infrared luminosity of the central source. Quasi-periodic flares in one or more Class II \methanol\/ maser lines  have been observed in over 20 objects \citep[periods$\sim$24-509 days, e.g.,][]{Goedhart2014,Szymczak15,Sugiyama18}; in one case, the 4.83~GHz \formaldehyde\/ maser also shows correlated flaring
\citep{Araya2010}. This periodic variability of \methanol\/ masers suggests a link to the mass gain process of high-mass stars.
Recently, two spectacular continuum outbursts in massive protostars have been accompanied by strong flaring of the \methanol\/ masers, S\,255IR~NIRS3 \citep{Caratti17} and NGC\,6334I-MM1 \citep{Hunter17}, the latter is shown in Figs.~1 and 2.  These outbursts are likely caused by a sudden increase in the accretion rate onto the central protostar. The event in S255 lasted only 2 years \citep{Liu18} while NGC\,6334I-MM1 continues in the flared state in millimeter continuum, methanol and water masers, and appears to be the counterpart in massive protostars to the FU~Ori phenomenon in low-mass protostars \citep{Francis19}.
\vspace{-5mm}
\begin{figure}[ht!]   
\centering
\begin{minipage}{0.6\textwidth}
\includegraphics[width=\linewidth]{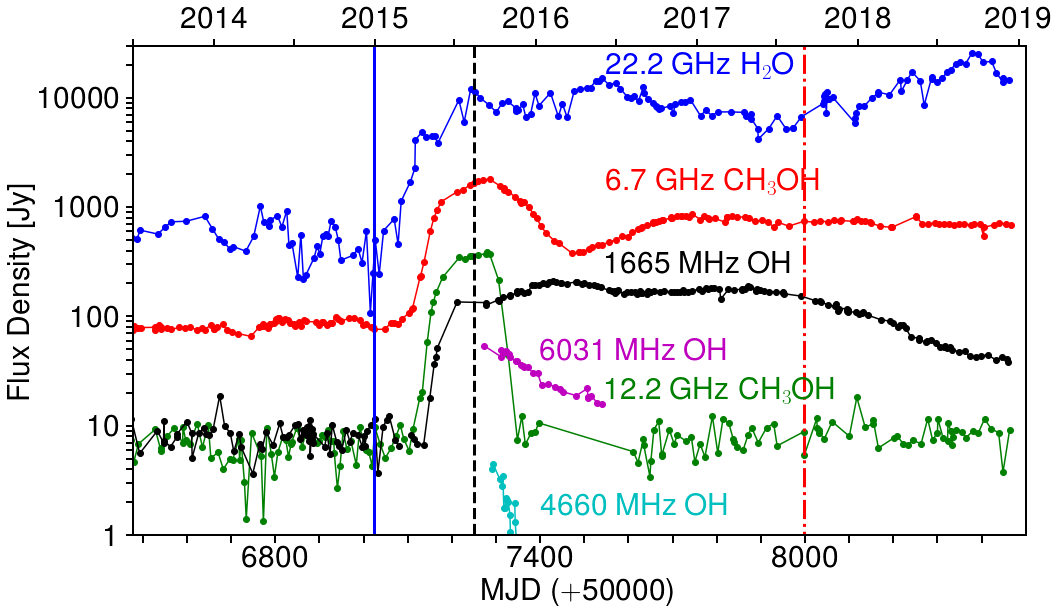}
\end{minipage}
\begin{minipage}{0.39\textwidth}
\includegraphics[width=\linewidth]{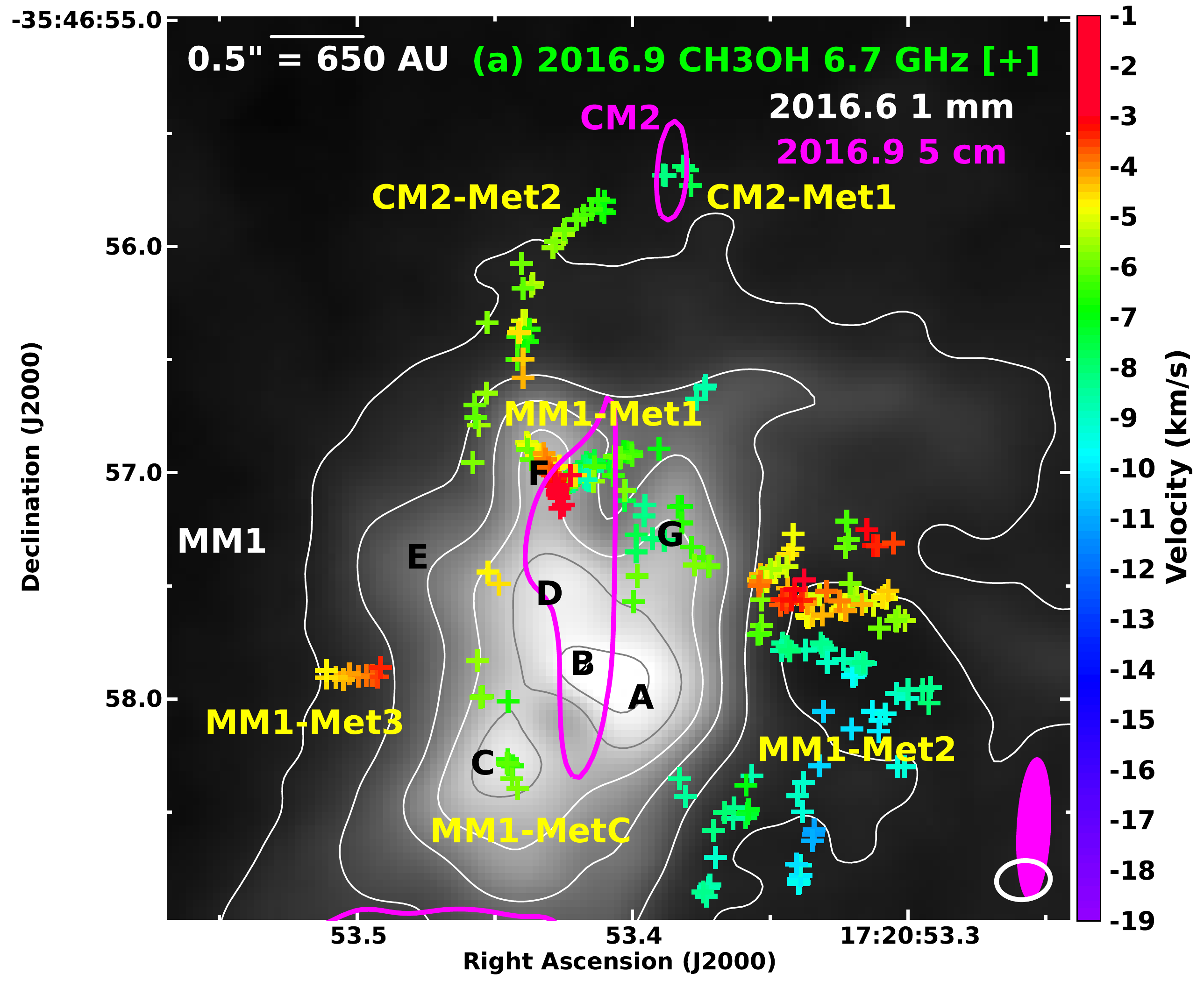}
\end{minipage}
\vspace{-5mm}
\caption{{\bf Left:} Single-dish light curve of six maser lines \citep{MacLeod18} which responded strongly to the ongoing accretion outburst in NGC\,6334I-MM1B, which also caused a quadrupling of the surrounding millimeter dust emission observed at ALMA in mid-2015 \citep{Hunter17}. {\bf Right:} VLA 6.7 GHz maser positions overlaid on ALMA 1~mm continuum image and VLA 5~cm contours \citep{Hunter18}. Prior to the outburst, MM1 never showed this maser in 3 decades of interferometric observations, but it is now the brightest source in this protocluster.  The spots trace dust cavities surrounding the central protostars B and D, where the density is not too high ($<10^8$~\cc) to quench the maser.}
\label{ngc6334}
\end{figure} 

\begin{figure}[ht!]   
\vspace{-4mm}
\includegraphics[width=0.98\linewidth]{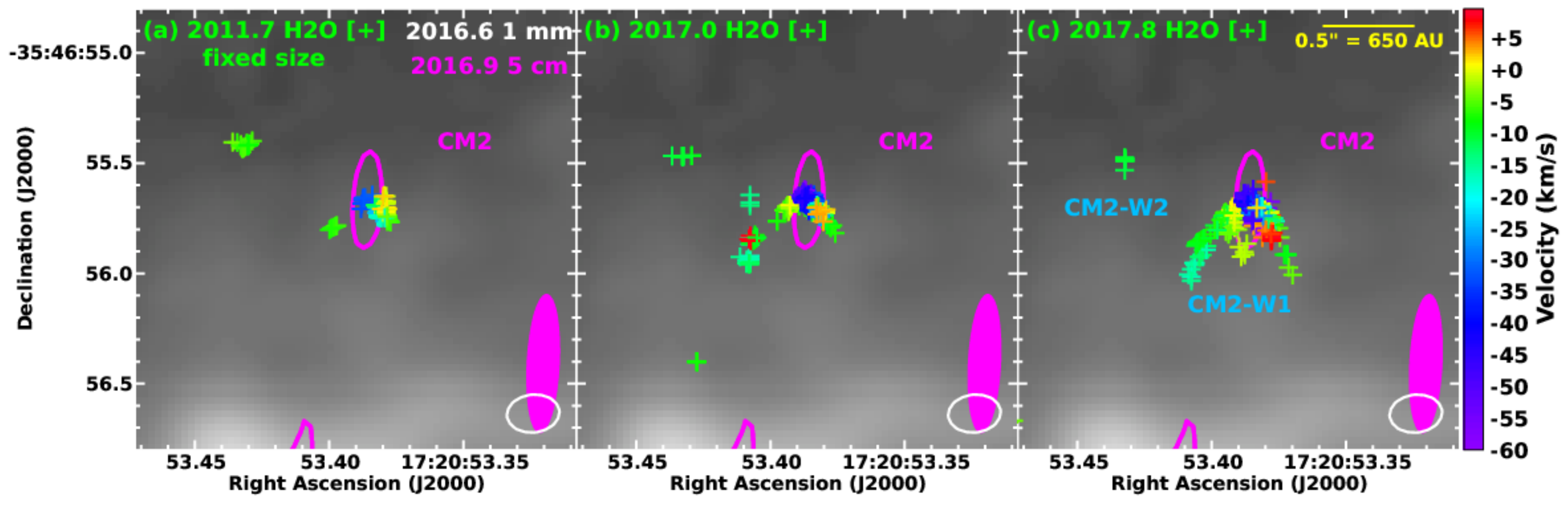}
\vspace{-5mm}
\caption{Pre-outburst (left) and post-outburst (center \& right) VLA water maser positions of the northern outflow lobe of NGC\,6334I-MM1. Radiation from the accretion outburst propagated 3000~au up the outflow cavity, igniting the maser flare which manifests in a parabolic bowshock whose apex coincides with synchrotron emission \citep{Brogan18}.}
\label{ngc6334water}
\vspace{-1mm}
\end{figure} 

These extraordinary events led to the formation of the international Maser Monitoring Organization (M2O), to promote single-dish monitoring and reporting of new maser flares to trigger interferometer studies while the accretion event is still underway.  
These accretion events can yield an increase in the thermal jet emission \citep{Cesaroni18}; or, in larger events, a decrease in the hypercompact HII region \citep{Brogan18b} due to the drop in UV radiation that results when the protostar swells, as recently predicted by numerical gravito-hydrodynamic simulations \citep{Meyer2019}.  Since both phenomena are powered by the protostar, the ability to perform simultaneous high-resolution observations of the continuum and the masers \citep{Hunter2018b} will enable direct measurements of correlations between them, yielding important constraints on the physics of the accretion mechanism. 

\begin{figure}[ht!]   
\vspace{-5mm}
\centering
\begin{minipage}{0.53\textwidth}
\includegraphics[width=0.97\textwidth]{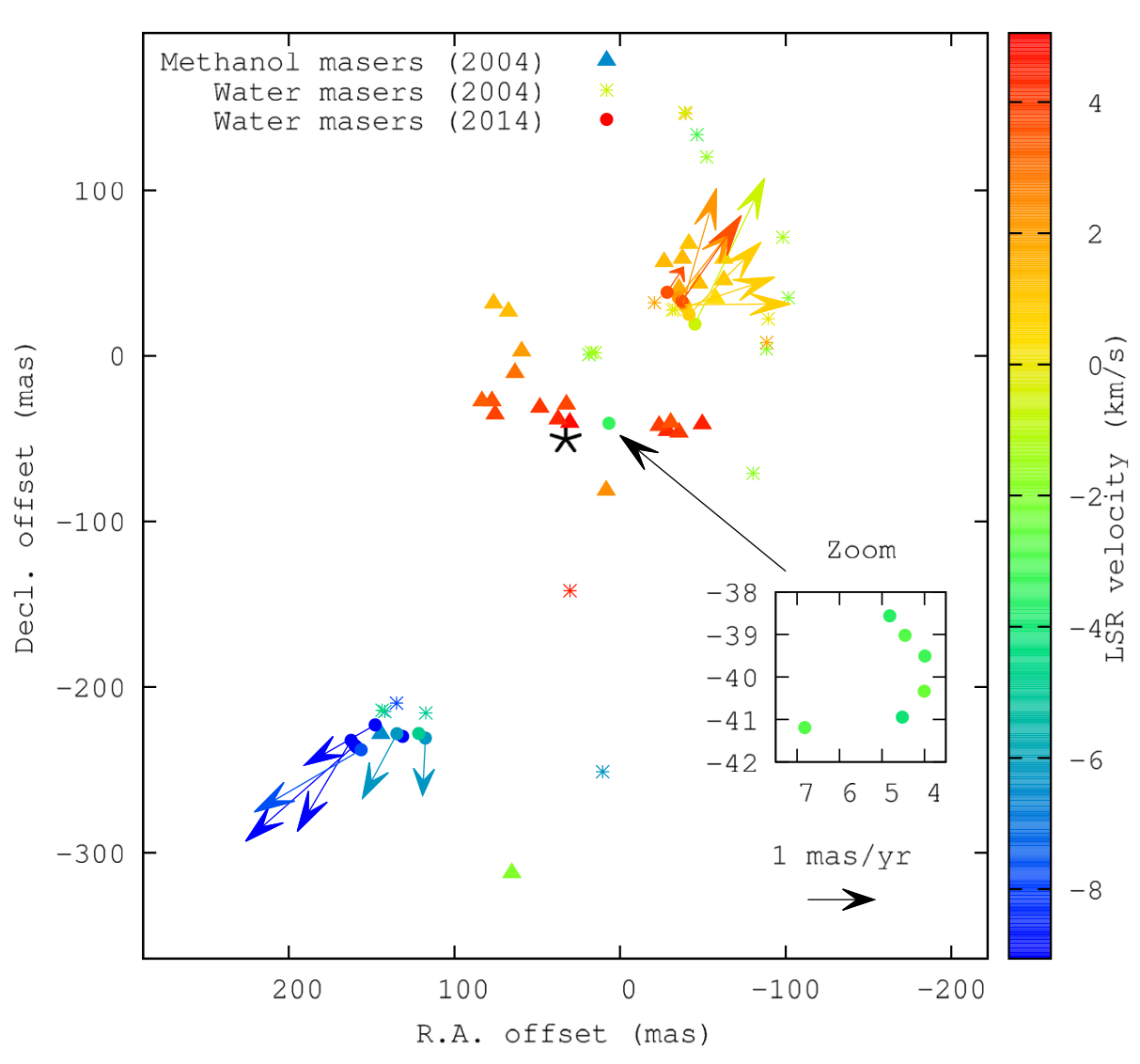}
\end{minipage}
\begin{minipage}{0.46\textwidth}
Figure 3: Central region of the dominant member (MM1) of the massive protocluster AFGL~5142, from \citet{Burns17}. Combined view
of 22-GHz water masers (filled circles) observed with VERA in 2010, 22-GHz water masers (colored asterisks) observed with the VLBA in 2004 \citep{Goddi06}
and 6.7-GHz methanol masers (triangles) observed with the EVN in 2004 \citep{Goddi07}. The inset shows the trajectory of feature A, moving
in a clockwise fashion. 
The black asterisk symbol
indicates the approximate origin of the episodic ejections.
\end{minipage}
\label{afgl5142}
\end{figure}

Similar to the Class II \methanol\/ maser lines, the 1.6~GHz ground state OH lines and several excited state OH lines (at 4.66~GHz, 4.75~GHz, 4.765~GHz, 6.030~GHz, and 6.035~GHz), are radiatively pumped  and will respond directly to luminosity outbursts (see Fig.~\ref{ngc6334}).  
The 22~GHz water maser line also traces gas close to massive and intermediate-mass protostars.  Although this line is primarily collisionally pumped, changes in the infrared radiation field can strongly affect the efficiency of the pump \citep{Deguchi81,Strelnitskii77}.   Water masers often span a broad velocity range, of several tens of \kms\/ about the systemic LSR velocity, particularly compared to methanol
masers ($\lesssim10$\,\kms). In some cases, water masers clearly arise from gas in the first few hundred au of the jet, such as in
Cepheus~A \citep[e.g.,][]{Torrelles2011,Chibueze12}, or in bow shocks somewhat further out \citep[e.g.,][]{Sanna2012,Burns16}. 
With continent-scale baselines, proper motion studies of these masers (see Fig.~\ref{afgl5142}) reveal the 3D velocities and orientations of collimated jets and/or wide-angle 
winds in the inner few 1000\,au from the central protostars \citep[e.g.,][]{Torrelles01,Torrelles2003,Torrelles2014,Moscadelli2007,Sanna2010,Burns17}.

\vspace{-6mm}
When these studies are combined with high-resolution radio continuum observations of radio thermal jets, they can allow us to quantify the outflow energetics
directly produced by the star formation process \citep[e.g.,][]{Moscadelli2016,Sanna2016}, as opposed to estimates of the molecular outflow energetics
that are attainable on scales greater than 0.1\,pc which suffer from source confusion. Long-term monitoring studies demonstrate that
water masers are also highly variable \citep[e.g.,][]{Felli07}. Since water masers are fundamentally produced in specific ranges of gas density and temperature within shocked gas layers,
they trace different types of coherent motions at different stages of protostellar evolution. A good example is W75\,N, where the 22~GHz masers and radio continuum show different spatial distributions around two distinct young stars  \citep{Torrelles2003,Carrasco2015}. The 22~GHz line is also unique in exhibiting the `superburst' phenomenon, in which brief
flares reach 10$^5$~Jy or more.  This has happened in only a few objects, including Orion~KL \citep[][]{Hirota14} and
G25.65+1.05 \citep{Lekht18}, but it has repeated in both, and appears to be due to interaction of the jet with high density clumps in the ambient
gas, within 3000~au of the central protostar. 
Finally, SiO maser emission from vibrationally-excited levels offers a powerful (but rare) probe of the innermost hot gas surrounding massive protostars.  For example, in one spectacular nearby case (Orion~KL), movies of the  v=1 and v=2 transitions of SiO J=1-0 at 43~GHz using the 
VLBA reveale a complicated structure of disk rotation and outflow in the inner 20-100~au \citep{Matthews2010}.  Additional massive protostars (at greater distances) have recently been detected in these lines \citep{Cordiner16,Ginsburg15,Zapata09}.  The increased sensitivity from new facilities  will yield further detections and enable detailed images of the inner accretion structures.

\vspace{-7mm}
\section{Sensitivity and resolution requirements}
\vspace{-4.5mm}

To fully exploit the investigative potential of masers,
baselines up to 9000~km are needed to reach resolutions of 1.5-0.15~mas at 5-50~GHz; at 4~kpc this translates to
5~au resolution in the 6.7~GHz methanol line, 1.4~au in the 22~GHz water line, and sub-au in the SiO lines.  At these scales, strong masers can be used 
to measure proper motions around the young protostar \citep{Sanna2010,Goddi2011}. With S/N$>$10, maser positions can be determined with an accuracy of 0.025~mas,
allowing proper motion measurements of 3~\kms\/ at 4~kpc in only two months. 
Unfortunately, the current $10\sigma$ sensitivity of the VLA and VLBA in 1-hr is only 0.1 and 0.3~Jy, respectively, in a 0.25~km/s channel. However, maser features can only be associated with physical structures when placed into context, so it is crucial to also detect significantly weaker masers to delineate kinematic structures like lines, rings, and bow shocks (Figs.~\ref{ngc6334water}, \ref{afgl5142}).  A factor of 10 improvement in sensitivity will provide this morphological context, and expand the number of sources for which proper motions are possible.
A survey of dozens of high-mass star-forming regions could be performed, providing baseline images to compare to future outbursts found by single dish maser monitoring.
Flexible, triggered scheduling will be important to rapidly image these flares in all maser lines to characterize the onset of the accretion burst, and ultimately, to understand the physical mechanism.



With the sensitivity required for the masers, the continuum emission from jets down to intermediate-mass protostars in the same cluster will be readily detectable.  In
nearby regions like Serpens (400~pc),
emission from clumps along the jet path typically does not exceed $\approx1$~mJy at 2-20~GHz \citep{RK2016}, which translates to only 6~$\mu$Jy for similar examples that populate more massive protoclusters at 4~kpc. In order to measure the spectral energy distribution (SED) of such an object, and distinguish free-free from synchrotron components, we would require a $6\sigma$ detection per GHz of bandwidth.  The current VLA requires 44~hr to reach $1~\mu$Jy rms at 3.6~cm alone. With an order of magnitude increase in effective collecting area, this time requirement would drop to $\lesssim0.5$~hr per band, meaning that a simultaneous maser/continuum multi-band survey of many fields would be feasible.  Also, with this sensitivity, chromospherically active young lower-mass T~Tauri stars, which have highly-variable faint (synchrotron) radio continuum emission, will be detected in the lower frequency bands, providing information about the low-mass population \citep{Forbrich17}. 
Photometry obtained from a sequence of continuum images, made with matched resolution of $0.05''$ from 5-100~GHz, will provide the SEDs of all the individual protostars on scales of 200~au $\times (d/4~{\rm kpc})$, giving an immediate census of the protocluster.  Synergistic images of thermal molecular line emission will enable kinematic comparisons with the maser lines, providing a powerful tool as recently demonstrated by comparing ALMA Band 10 images of HDO with VLA water masers \citep{McGuire2018}.  
%
{\it In summary, larger radio interferometers, like the planned SKA (at longer wavelengths) and ngVLA (at shorter wavelengths), are required to open a new page in our understanding of how the most massive  stars form, which remains a major open question of modern astrophysics.}



\section*{Acknowledgements} The National Radio Astronomy Observatory is a facility of the National Science Foundation operated under agreement by Associated Universities, Inc.
The Maser Monitoring Organization (M2O) is a voluntary organization hosted by the South African Radio Astronomy Observatory (SARAO) and can currently be reached at {\tt https://m2o.hartrao.ac.za}, with a future migration planned to {\tt https://MaserMonitoring.org}

\end{document}